\documentclass{article}

\usepackage{microtype}
\usepackage{graphicx}
\usepackage{subcaption}
\usepackage{booktabs} 
\usepackage{url}
\usepackage{graphicx}
\usepackage{makecell} 
\usepackage[table,xcdraw]{xcolor}
\usepackage{multirow}
\usepackage[ruled,vlined]{algorithm2e}
\usepackage{placeins}
\usepackage{dblfloatfix}
\usepackage{float}

\usepackage{hyperref}

\usepackage[accepted]{icml2026}
\usepackage{amsmath}
\usepackage{amssymb}
\usepackage{mathtools}
\usepackage{amsthm}

\usepackage[capitalize,noabbrev]{cleveref}

\theoremstyle{plain}

\theoremstyle{definition}

\theoremstyle{remark}

\usepackage[textsize=tiny]{todonotes}

\icmltitlerunning{Beyond Generalist LLMs: Specialist Agentic Systems for Structured Code Workflow Execution}

\begin{document}

\twocolumn[
  \icmltitle{Beyond Generalist LLMs: Specialist Agentic Systems for Structured Code  \\
    Workflow Execution}



  \icmlsetsymbol{equal}{*}

  \begin{icmlauthorlist}
   
    \icmlauthor{Harris Borman}{equal,comp}
    \icmlauthor{Herman Wandabwa}{equal,comp}
    \icmlauthor{Fusun Yu}{comp}
    \icmlauthor{Sandeepa Kannangara}{comp}
    \icmlauthor{Justin Liu}{comp}
    \icmlauthor{Anna Leontjeva}{comp}
    \icmlauthor{Ritchie Ng}{comp}
  \end{icmlauthorlist}

  \icmlaffiliation{comp}{Commonwealth Bank of Australia, Sydney, Australia}

  \icmlcorrespondingauthor{Harris Borman}{harris.borman@cba.com.au}
  \icmlcorrespondingauthor{Herman Wandabwa}{herman.wandabwa@cba.com.au}

  \icmlkeywords{Machine Learning, ICML}

  \vskip 0.3in
]



\printAffiliationsAndNotice{}  

\begin{abstract}

Large Language Models (LLMs) have accelerated the adoption of software development agents, now widely available as Integrated Development Environment (IDE) extensions and standalone applications. While these agents are typically general-purpose, it remains unclear whether specialist agents justify their additional development effort. We investigate this question in the context of business process automation, focusing on the transformation of Business Process Model and Notation (BPMN) diagrams into executable agentic workflows. Since BPMN specifies explicit control-flow semantics, we focus on deterministic workflows in which a fixed process model and inputs uniquely determine the executed path. We introduce a specialist workflow for this task and compare it against generalist agents such as Roo and Cline. Our results show that the specialist solution produces agents that outperform generalist baselines by approximately 9--20 percentage points in tool-use exactness, 2--4$\times$ in penalty-adjusted latency, and 3$\times$ fewer tool-call errors, while reducing generation token cost by over 95\% and eliminating repair iterations. We also find that generalist agents generate code inconsistently in both functionality and quality, limiting their suitability for industrial settings where reliability and maintainability are essential.

\end{abstract}


\section{Introduction}
\label{sec_intro}

The emergence of LLMs has accelerated the rise of autonomous software agents \cite{ferrag2025llmreasoningautonomousai}. These AI-driven agents now appear as IDE extensions and stand-alone no-code assistants, enabling even non-programmers to build simple applications within minutes \cite{10.1145/3712003}. Most operate as generalists, leveraging foundational LLMs to perform a wide range of tasks. For example, an open-source coding assistant like Roo Code can plan, write, and debug code across domains directly in a developer's editor \cite{sapkota2025vibecodingvsagentic}. Similarly, multi-agent frameworks such as FLOW, AFLOW, AutoGen, or MetaGPT coordinate several LLM agents with predefined roles to solve complex problems in a general way \cite{niu2025flow,zhang2025aflow,wu2024autogen,hong2024metagpt}. These systems have been applied to tasks ranging from web browsing and data analysis to game design and UI creation~\cite{fourney2024magenticonegeneralistmultiagentsolving}. Prior work has largely focused on what we define as “generalist systems”, a system of agents that are able to complete a wide range of tasks, with architectures suited to free exploration of various ideas to complete a task in an unspecified manner \cite{Sapkota_2026}. This adaptability has driven adoption, including IDE extensions such as Roo and Cline \cite{sapkota2025vibecodingvsagentic,Cline}. These systems are useful because they can create a complete system from a single prompt with minimal user intervention, allowing users with limited technical knowledge to build a functioning system from a simple idea \cite{Sapkota_2026}.

However, these systems have drawbacks. Their generalist design often demands extensive planning and increases token and cost overhead, especially during rapid or repeated development. They may also lack awareness of company-specific best practices, such as style guides or preferred methods. While experienced users can impose constraints and refine outputs, this remains imperfect and does not guarantee consistency across many generations, as discussed in Section \ref{sec_methodology}. This inconsistency can increase technology debt and complicate future updates and maintenance \cite{10.1145/3756681.3756976}.

An alternative paradigm is what we define as a “specialist system”. These systems use a well-defined and constrained agentic workflow designed for a specific class of tasks. Rather than asking an LLM to generate solutions from scratch, specialist systems encode expert knowledge into a templated workflow. Within this scaffold, the LLM performs localised reasoning, making small, context-sensitive adjustments within predefined steps rather than constructing the full solution from first principles.

These systems usually require greater upfront effort because the workflow must be manually designed and validated. This makes them less suitable for short-lived use cases such as demos or proof-of-concept experiments. However, they are better suited to repeated or large-scale deployment. By producing more consistent outputs, they can reduce technical debt and simplify debugging and integration with external systems. Their structured design also enables tighter context management, allowing developers to control what information is exposed to the LLM, reduce unnecessary token usage, and improve performance at scale.

Recent studies show that although LLMs are increasingly equipped with extended context windows, their utilisation of this capacity remains uneven \cite{an2025why}. Empirical evidence also suggests that performance tends to degrade as more of the context window is consumed \cite{modarressi2025nolimalongcontextevaluationliteral,laban2025llms}. To address this, we propose a context management strategy that restricts the active context to the minimum information required for each subtask. In generalist systems, this is difficult because it requires prior knowledge of the information needed for each subtask during workflow construction, and such designs are often task-specific and do not transfer easily across domains. Restricting context in this way can reduce redundancy and improve model performance relative to systems that retain excess context indiscriminately, as shown in Section~\ref{sec_results}. Given that business processes can scale well beyond the complexity of our benchmark workflows, effective context management is critical for maintaining performance in large-scale deployments.

Efficient context management also offers potential cost benefits. This is especially important in business environments where standard operating procedures may evolve frequently during development and post-deployment phases and agentic workflow generation tools may be executed repeatedly for the same task. Under these conditions, reducing token volume can yield substantial savings \cite{mei2025surveycontextengineeringlarge,laban2025llms}. While the per-instance reduction may seem small, the cumulative impact at scale can be significant. This creates value for specialist workflows that use manually constructed context management to minimise the information exposed to the LLM in a targeted way while maintaining performance and reducing cost.


To operationalise these specialist workflows and the scoped context they enable, we adopt a modular code generation approach in which the source of decomposition is external to the model. Rather than discovering plans through prompts, roles, or library modules as in prior multiagent systems (e.g., AutoGen, MetaGPT, DSPy, SWE-agent)~\citep{wu2024autogen,hong2024metagpt,khattab2024dspy,yang2024sweagentagentcomputerinterfacesenable}, we compile an industry standard BPMN 2.0 process model into a ReAct style control graph. Each BPMN node provides a typed tool contract and a node local policy, while the control plane enforces branches and joins, scopes context per node, and applies contract derived runtime validation with targeted retries. This specifications-to-agent compilation yields an auditable workflow structure, minimises unnecessary context exposure, and avoids rediscovering plans at runtime. The design is motivated by three principles: constrained execution via BPMN-derived control flow, targeted context management that limits irrelevant information exposure, and modular decomposition that supports more reliable tool-level reasoning. Against this background, our study examines whether a specialist BPMN-grounded workflow can offer practical advantages over more generalist agentic systems. The following sections describe how this design is instantiated using BPMNs, outline the experimental methodology and benchmark design, and present results comparing specialist and generalist agentic workflows.

\subsection{Business Processes}
\label{sec_business_processes}

To ground our study, we focus on business process automation and adopt BPMN as the structural representation of workflows. BPMN is widely used in enterprise modelling and provides a practical interface through which non-technical users can specify process logic in a standardised form~\citep{kopke2024bpmnchatbot,nour2025nala2bpmn,moraes2025automating,berti2024pmllm}. This makes it a suitable domain for studying how LLM-based agentic systems can transform human-authored process specifications into executable workflows.

In this work, we consider the task of converting BPMN-defined workflows into operational agentic pipelines. This setting is appropriate for our comparison between specialist and generalist systems because BPMN provides an explicit and auditable representation of process structure, allowing us to assess how effectively each approach preserves intended workflow logic while supporting executable automation.

\section{Literature Review}
\label{sec_lit_review}

\subsection{BPMN and Traditional Workflow Execution}
\label{sec_bpmn_and_traditional_workflow_execution}

Business Process Model and Notation (BPMN) is a standard for modelling structured business workflows. Its graphical notation is both human-readable and machine-executable, which enables organisations to define, automate, and monitor processes effectively~\citep{white2004introduction,chinosi2012bpmn,dumas2018fundamentals}. Traditional BPMN engines operate in deterministic, rule-based settings involving human tasks, service calls, and decision gateways~\citep{weske2019business}. While robust in predictable environments, they provide limited support for dynamic or context-sensitive decision-making~\citep{van2020challenges}. Prior work has tried to address this through agent-based automation~\citep{wooldridge1995intelligent}, adaptive workflow systems~\citep{reichert2012enabling}, context-aware frameworks~\citep{rosemann2008contextualisation}, and decision-centric models~\citep{batoulis2015extracting}. BPMN extensions have also been proposed to support more adaptive workflows~\citep{braun2014bpmn4cp}, but execution remains largely constrained by static semantics~\citep{mendling2018blockchains}. As a result, traditional BPMN execution still struggles with unstructured data, runtime variability, and ambiguous decision logic~\citep{marrella2019automated}.

\subsection{LLMs and Agentic Workflows}
\label{sec_llms_and_agentic_workflows}

Recent advances in Large Language Models (LLMs) have expanded their role from passive predictors to agents capable of reasoning, planning, and executing tasks from natural language instructions~\citep{wei2022chain}. Agentic workflows build on this by combining reasoning with tool use to enable autonomous task completion with limited human oversight~\citep{schick2023toolformer}. Architectures such as ReAct~\citep{yao2023react}, Tree-of-Thoughts~\citep{yao2023treeofthoughts}, and PAL~\citep{gao2023pal} illustrate this shift toward adaptive task coordination and decision-making~\citep{yang2023foundation}. Research has also explored external memory, planning modules, and tool integration~\citep{wu2024autogen}. Systems such as AFLOW~\citep{zhang2025aflow} and MaAS~\citep{zhang2025multiagent} further show the promise of agentic AI for complex tasks such as retrieval, analysis, and decision-making~\citep{singhal2023large}. However, these systems also raise challenges around traceability, control, and integration with structured workflow representations such as BPMN~\citep{mialon2023augmented,deng2023mind2web}. Most remain general-purpose frameworks, with less attention given to specialist workflows for narrowly defined structured automation tasks.

\subsection{BPMN and LLM Integration}
\label{sec_bpmn_and_llm_integration}

Initial work on BPMN and LLM integration has focused mainly on modelling rather than execution. Representative directions include generating BPMN diagrams from text, conversational refinement of process models, and workflow mining~\citep{kopke2024bpmnchatbot,nour2025nala2bpmn,moraes2025automating,berti2024pmllm}. These approaches improve accessibility for non-experts, but they generally treat BPMN as a static artefact rather than a basis for executable agentic behaviour.

More recent work has moved toward agentic automation, where LLMs synthesise workflows and execute tasks across tools and APIs~\citep{jain2024smartflow,zeng2024flowmind,ye2023proagent}. Despite this progress, a key gap remains in combining BPMN's formal process structure with the flexibility of LLM-driven agents. In particular, reliably executing BPMN-defined workflows while preserving semantic rigour and handling unstructured inputs remains an open challenge.

\section{Methodology}
\label{sec_methodology}

To evaluate the efficiency and performance gains of a specialist agent, we designed one for converting BPMN-specified workflows into ReAct agents. We evaluated each system using metrics that capture both the agent generation process and the performance of the generated agents, enabling comparison of system efficiency and output quality.

\subsection{Workflow Selection}
\label{sec_workflow_selection}

We evaluated our approach using ten deterministic workflows of varying complexity. The workflows were manually constructed by the authors to reflect business process automation tasks across multiple domains, including e-commerce, cost optimisation, risk, and information retrieval. Although this workflow set was not derived from an established benchmark, its construction was informed by foundational business process modelling research. Specifically, we drew on research on business process families and variants \cite{laRosa2017variability,delgado2022bpmnFamilies}, process model quality and comprehension \cite{mendling2010seven,figl2017comprehension}, and representative BPMN model generation \cite{skouradaki2016representative} to guide workflow diversity and structural complexity.


The workflows were manually designed because we are not aware of an established benchmark for evaluating the conversion of BPMN-style workflows into executable agentic systems. Existing workflow datasets and benchmark resources generally focus on process discovery, conformance checking, event logs, or model analysis rather than end-to-end evaluation of workflow-to-agent translation and execution fidelity \cite{vanDerAalst2022processMining,tfpmEventLogs,burattin2016plg2}. Table~\ref{tab:workflow_summary} in the Appendix summarises each workflow, including the number of nodes (tasks, gateways, and events) and edges (sequence flows), which we use as indicators of structural complexity.


The workflows were selected according to two criteria. First, they span a broad range of structural complexity, from 9 to 52 nodes, allowing evaluation across processes of different sizes and control-flow depth. Second, all workflows are deterministic, in the sense that each execution path is governed by predefined labels and conditions and yields a directly specifiable expected outcome. This design lets us isolate the core capability studied in this paper and evaluate it systematically through exhaustive path coverage. We acknowledge that author-constructed workflows may introduce design bias and that restricting the evaluation to deterministic processes limits generalisability, particularly to settings involving ambiguity, stochasticity, or open-ended human decision making. The results should therefore be interpreted as evidence for structured, deterministic process settings rather than as a claim of universal representativeness.

\subsection{Agentic System Design}
\label{sec_agentic_system_design}

Although these workflows could be executed using a fixed Directed Acyclic Graph (DAG), and loops can be achieved via LangGraph, we instead evaluate the ability of our system to construct a ReAct-based solution \citep{yao2023react}. A ReAct agent conventionally plans actions and calls tools through an interactive loop rather than following a strictly predefined execution path. We chose this formulation for three reasons:
\begin{enumerate}
    \item \textbf{Adaptability beyond fixed execution:} ReAct agents can generalise across a broader range of tasks and respond more flexibly to changing execution conditions. In practical settings such as customer service chatbots, workflows may require extracting information from user messages, calling multiple tools, handling corrected inputs, or revising earlier decisions. These scenarios benefit from an agent that can re-plan during execution rather than follow a rigid script \cite{LEOCADIO20241222}.

    \item \textbf{Scalability to realistic workflows:} Compared with static DAG execution, agent-based control is better suited to workflows that extend beyond fixed linear paths, particularly when user inputs, state changes, or partial task completion require dynamic coordination across steps.

    \item \textbf{Ease of authoring:} The logic of a ReAct agent is expressed in natural language through prompt instructions, which can make workflows easier for non-technical subject matter experts to understand and modify. Minor changes to behaviour can therefore be made by editing instructions rather than altering code, which can speed up iteration and development.
\end{enumerate}

\subsection{Proposed System}
\label{sec_proposed_system}

\begin{figure*}[!ht]
\centering
\includegraphics[width=1.0\textwidth]{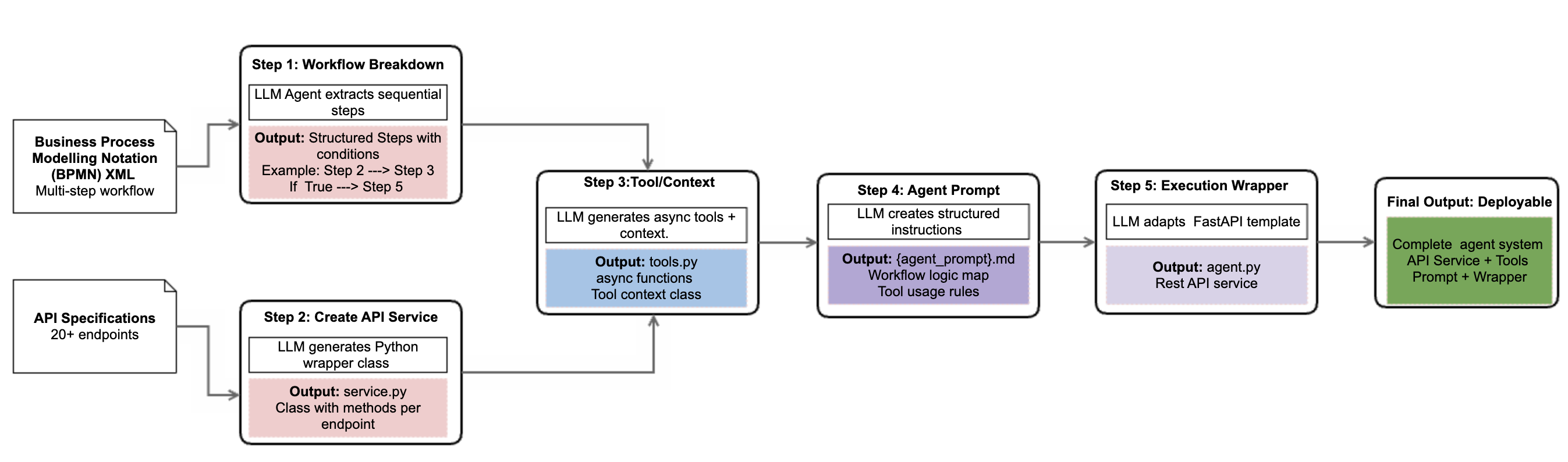}
\caption{Proposed System Architecture}
\label{fig:proposed_system}
\end{figure*}


Our proposed system (Figure \ref{fig:proposed_system}) takes a BPMN-defined workflow and API specifications and automatically generates a working ReAct-style agent through the following steps:

\begin{enumerate}

    \item \textbf{Workflow Parsing:} The BPMN diagram is parsed into discrete steps, identifying tasks, decision nodes, and required API calls. This yields a structured representation of the workflow logic (including branches and conditions) that the language model can reason about.

    \item \textbf{API Service Generation:} From the provided API specification, the system generates a reusable client module that wraps external calls and abstracts low-level details, reducing tool implementation complexity and code duplication.

    \item \textbf{Tool/Context Creation:} Using the parsed workflow steps and the API service, the system generates code for each tool corresponding to a workflow node.

    \item \textbf{Iterative Refinement (Agent Self-Verification):} 
    Generated tool code is executed and validated against expected behaviour. On failure, the system enters a refinement loop: the language model analyses the error and revises the implementation. This loop continues until execution succeeds or successive iterations cease to make substantive progress. Each component is thereby either validated or identified as unresolved.

    \item \textbf{Agent Assembly and Deployment:} Once all tools are verified, the system composes a natural language prompt encoding the workflow logic and generates a main function that instantiates the ReAct agent behind a FastAPI service. The resulting system—prompts, tools, and API endpoints—is then ready for end-to-end testing.

\end{enumerate}
 
\section{Evaluations and  Results}
\label{sec_evaluation}

\subsection{Baseline Systems and Experimental Setup}
\label{sec_baseline_systems_and_experimental_setup}

We evaluated our system against two automated coding agents, \textbf{Roo} and \textbf{Cline}, available as Visual Studio Code extensions~\cite{vscode2025}. All systems received identical BPMN workflows, API specifications, and backend LLM. Roo and Cline were allowed to operate unconstrained with their default approaches, as preliminary experiments showed that imposing additional design constraints degraded output quality. The base prompt is shown in Figure~\ref{fig:base_prompt}, where the \texttt{<BPMN>} tags contain the raw BPMN 2.0 XML for each workflow.

After each system declared completion, we conducted end-to-end verification on sample inputs covering all workflow branches. If an agent failed, we fed the error back and allowed iterative self-repair until the issue was resolved or the system could no longer make progress. Our system's agents passed all end-to-end tests on the first attempt without manual intervention (Section \ref{sec_agent_generation_results}), unlike those generated by Roo and Cline.

We also evaluated AutoGen, MetaGPT, and FLOW~\cite{wu2024autogen,hong2024metagpt,niu2025flow}, but none produced functional solutions for any workflow. MetaGPT's outputs were typically incomplete (e.g., generating only prompts and tools but not the full agent). AutoGen, even with its GraphFlow controller, failed to autonomously generate the required files without manual agent wiring. FLOW produced high-level task decompositions but not complete, executable code. As none completed even a single workflow end-to-end, we excluded them from the comparative evaluation, focusing on Roo and Cline as the only baselines that successfully completed the tasks.

\subsection{Evaluation Metrics}
\label{sec_evaluation_metrics}

We evaluate each system across two tasks: 
\begin{enumerate}
    \item the agent generation process, and
    \item the functional performance of the generated agents.
\end{enumerate}


For each system under comparison (Roo, Cline and our system), we generated at least 10 agents that successfully compiled and executed according to the BPMN-defined specifications for each of the ten workflows. Each of the 10 agents was then evaluated on a dataset comprising all possible combinations of control-flow flags, yielding comprehensive test coverage across all execution paths. As summarised in Table~\ref{tab:workflow_summary}, the workflows vary in structural complexity from 9 nodes and 10 edges (Weather) to 52 nodes and 60 edges (Cart), spanning diverse business domains. This diversity ensures that our evaluation captures a wide range of real-world scenarios, providing a robust testbed for assessing the capabilities of each system in generating functional agents from BPMN specifications.

Achieving the target of 10~successful agents per workflow per system required a total of 371~generation attempts across all three systems, of which 300 succeeded and 71 failed outright. Our system and Cline each needed 117~attempts to obtain their 100~successful agents (85.5\% success rate), while Roo required 137~attempts (73.0\% success rate). The 300~successful agents were collectively evaluated on 30{,}543~individual test cases (approximately 102 per run), covering all possible control-flow paths in each workflow. A generation attempt was classified as a \emph{failure} when the coding agent was unable to produce a functional agent at all, for example due to API errors, empty output, or code that could not be executed end-to-end. This is distinct from \emph{repair iterations} (discussed in Section~\ref{sec_agent_generation_evaluation}), where the agent initially produced failing code but self-corrected after receiving error feedback. Failure rates varied across workflows: simpler workflows such as Social, Spam, and News were completed without any failures by all three systems, whereas Weather (46.4\% failure rate; 56~attempts for 30~successes) and Tournament (42.3\%; 52~attempts) proved most challenging. To assess cross-run consistency, we computed the coefficient of variation (CV) of the Tool-Use Exactness score across the 10~independent runs per workflow: our system exhibited the lowest variability (CV\,=\,0.63), followed by Cline (0.86) and Roo (1.05), indicating that our approach produces more consistent agents across independent generations.


\subsubsection{Agent Generation Evaluation}
\label{sec_agent_generation_evaluation}

The efficiency of the agent generation process was assessed using two primary metrics: the number of \emph{repair iterations} and \emph{token usage}.

\begin{enumerate}
    \item \textbf{Repair Iterations:} We measured the number of refinement loops required to produce a valid agent. For our system, this corresponds to the number of tool-refinement cycles needed to achieve successful execution. For Roo and Cline, this includes code errors and errors in the FastAPI service generated to invoke the agent. Notably, our system’s templated approach to developing a FastAPI service that wraps the agent resulted in no errors in this category. We report this as the average number of repair iterations required to complete a single successful generation. Repairs performed in generations where the task was not successfully completed were not included.
    
    \item \textbf{Token Usage and Cost:} We tracked the total number of tokens consumed from the initial invocation through the successful generation of a complete agent, recording input, output, and total token counts. Our system’s token usage was measured using Langfuse tracing, while Roo and Cline reported their own token consumption.
\end{enumerate}

\subsubsection{Agent Run Evaluations}
\label{sec_agent_run_evaluations}

Each generated agent was evaluated on the following metrics using GPT-4.1 for consistency.

\begin{enumerate}

  \item \textbf{Process Adherence:} Assessed via an LLM-as-a-judge evaluation. The judge received (i) the workflow specification defining the expected tool-call sequence, (ii) the input request, and (iii) the agent's execution trace. Each tool call was categorised as \emph{correct}, \emph{missed} (required but not called), \emph{excess} (called but not required), or \emph{out-of-sequence}. The judge returned a binary adherence verdict (\texttt{deviation: true/false}) with a step-by-step reasoning trace.

  \item \textbf{Tool-Use Exactness (TUE) Score:} TUE is a binary per-run indicator equal to 1 when the agent invokes exactly the prescribed tools with zero missed, excess, or out-of-sequence calls, and 0 otherwise. The aggregate TUE score is the percentage of runs achieving a perfect score.

  \item \textbf{Penalty-Adjusted Latency:} We define latency as the time required per unit of \emph{net-correct workflow progress}, penalising runs that spend time on incorrect tool usage. Let $T$ denote total execution time, $C$ the number of correct tool calls, and $M$, $E$, $O$ the counts of missed, excess, and out-of-sequence calls. The effective steps are
\begin{equation}
P = \max(\varepsilon,\; C - (M + E + O)),
\end{equation}
where $\varepsilon=1$ avoids division by zero. The penalty-adjusted latency is then:
\begin{equation}
\mathrm{Latency} = \frac{T}{P}
\label{eq:latency}
\end{equation}
Lower values indicate faster, more accurate execution.

\end{enumerate}

\subsection{Results}
\label{sec_results}

\subsubsection{Agent Generation Results}
\label{sec_agent_generation_results}

\begin{figure*}[!ht]
  \centering
  \includegraphics[width=\textwidth]{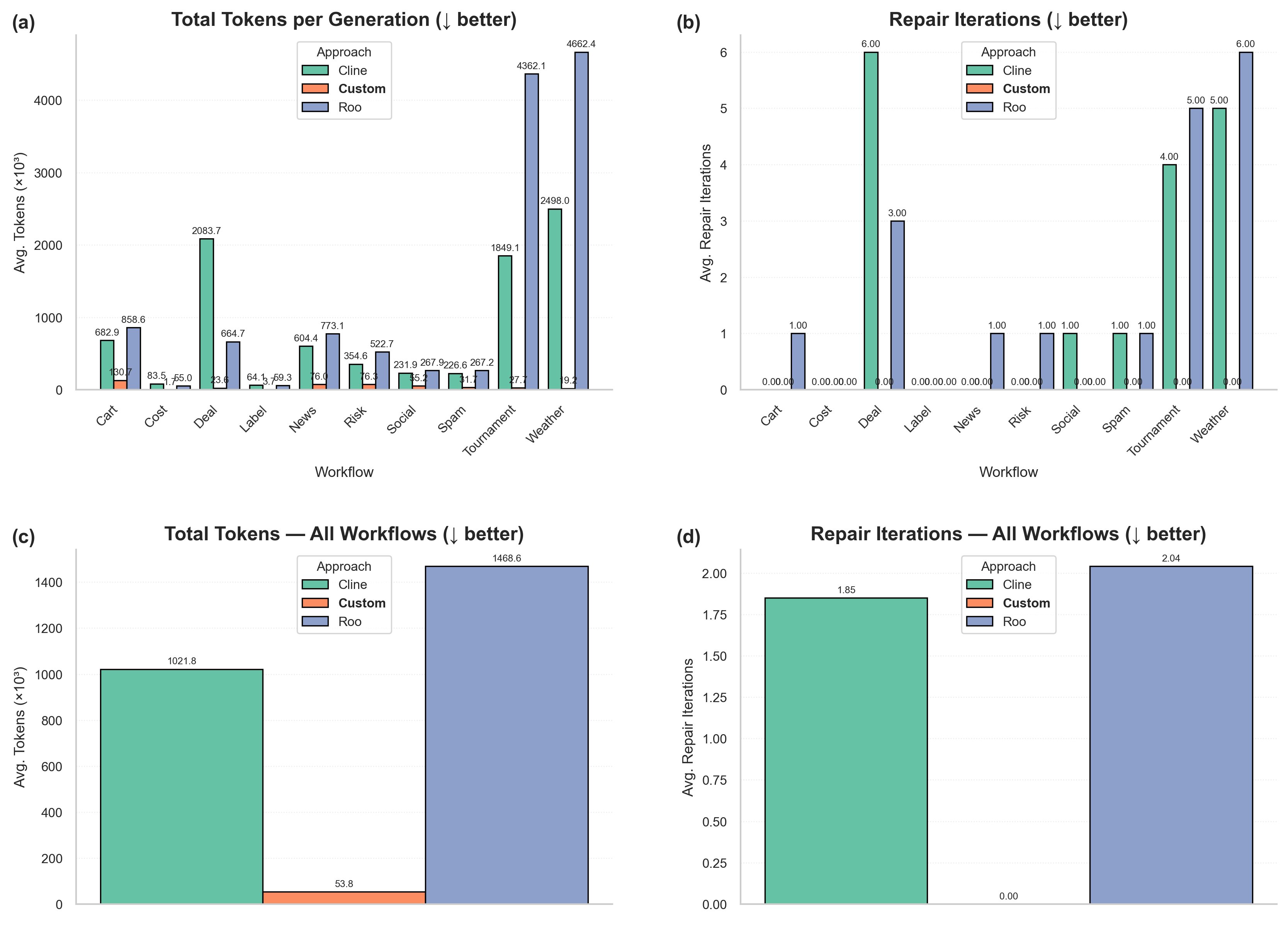}
  \caption{Agent Generation Metrics: Repair Iterations, Token Usage (Input and Output) for the Specialist System(Custom), Roo, and Cline.}
  \label{fig:agent_generation_metrics}
\end{figure*}

\begin{enumerate}
\item \textbf{Repair Iterations:} As shown in Figure~\ref{fig:agent_generation_metrics}, the specialist system required zero repair iterations across all successful generations, compared to averages of 2.08 for Roo and 1.89 for Cline, reflecting their reliance on iterative refinement loops to achieve functional outputs.

\item \textbf{Token Usage and Cost:} The specialist system averaged 49.24k input and 5.82k output tokens per agent, versus 1{,}484.46k/17.40k for Roo and
1{,}029.27k/15.36k for Cline (a 2{,}915\%/199\% and 1{,}990\%/164\% increase respectively), as shown in Figure~\ref{fig:agent_generation_metrics}. Combined with zero repair iterations, the specialist system produces a correct agent in a single pass at $\approx$55k total tokens, compared to over 1{,}500k for Roo and 1{,}044k for Cline before accounting for additional tokens consumed in repair cycles. This order-of-magnitude difference makes specialist pipelines substantially more cost-effective at scale.

\end{enumerate}

\subsubsection{Agent Run Evaluation Results}
\label{sec_agent_run_results}

\begin{itemize}

\item \textbf{Tool-Use Exactness (TUE):}
Our specialist system achieved the highest TUE score at 57.69\%, compared to 48.62\% for Cline (+9.1\,pp) and 38.11\% for Roo (+19.6\,pp) (Figure~\ref{fig:appendix:key_metrics}(a)). The per-workflow breakdown in Figure~\ref{fig:agent_run_evaluation_metrics}(a) confirms this advantage holds consistently across all ten workflows regardless of complexity.

\item \textbf{Tool-Call Errors:}
The specialist system averaged 1.27 total errors per run versus 3.19 for Cline and 3.22 for Roo (both 2.5$\times$ higher; Figure~\ref{fig:appendix:key_metrics}(b)). The dominant error mode was missed tool calls (0.86 specialist, 2.05 Cline, 1.76 Roo), reflecting generalist agents' tendency to return an output once sufficient information was gathered rather than completing the full prescribed sequence. Excess calls were less frequent but still elevated (0.95/Cline, 1.13/Roo vs.\ 0.37/specialist), and out-of-sequence calls were low across all approaches ($\leq$0.33). These patterns are consistent across workflows (Figure~\ref{fig:agent_run_evaluation_metrics}(b)).

\item \textbf{Process Adherence:}
The specialist system achieved the highest adherence rate at 54.68\%, versus 51.57\% for Cline (+3.1\,pp) and 42.43\% for Roo (+12.2\,pp) (Figure~\ref{fig:appendix:key_metrics}(d)). The gap widens on more complex workflows (Figure~\ref{fig:agent_run_evaluation_metrics}(d)), suggesting that generalist systems' tendency to omit tool calls compounds into lower process conformance as workflow complexity increases.

\item \textbf{Penalty-Adjusted Latency:}
The specialist system achieved the lowest latency at 2.49\,s per effective step, versus 6.59 for Cline (2.6$\times$) and 9.08 for Roo (3.6$\times$) (Figures~\ref{fig:appendix:key_metrics}(c), \ref{fig:agent_run_evaluation_metrics}(c)). The elevated generalist latency reflects both longer raw execution times and a higher incidence of tool-call errors, both of which inflate the metric.

\end{itemize}

Taken together, the specialist system outperforms both baselines on all four metrics, demonstrating the benefits of structured, specification-driven agent generation over general-purpose coding frameworks. These gains likely stem from the design principles outlined earlier in the paper: constrained execution via externally specified workflow structure, targeted context management, and modular decomposition. Together, these reduce search space and planning overhead, which in turn contributes to fewer tool-call errors and more consistent execution across runs.

\section{Conclusion}
\label{sec_conclusion}

We presented a specialist agentic system for converting BPMN-defined workflows into executable ReAct agents and evaluated it against Cline and Roo across ten workflows under the same foundation model. The specialist system outperformed both baselines on all four evaluation metrics, achieving higher tool-use exactness and process adherence, substantially lower penalty-adjusted latency, and fewer tool-call errors. It also reduced token consumption by more than 95\% relative to the generalist baselines and required no repair iterations. These gains stem from decomposing each workflow into modular prompt, tool, and orchestration components, enabling constrained execution and targeted context management rather than iterative trial-and-error generation.

These improvements are particularly valuable in enterprise settings, where reliability, maintainability, and cost efficiency are critical for large-scale deployment. By externalising workflow structure and reducing unnecessary context exposure, specialist systems can offer a more predictable and operationally manageable alternative to general-purpose agentic coding assistants for structured automation tasks.

The evaluation is scoped to deterministic workflows under realistic low-configuration usage. Although prompt engineering or iterative tuning could improve generalist performance, doing so would require additional user effort and expertise. Future work will examine more heavily optimised baseline configurations and conduct component-level ablations to determine how each structural element contributes to performance, and whether these components can be selectively incorporated into generalist assistants.


\section{GenAI Usage Disclosure}
\label{sec:genai_usage_disclosure}

In line with our experimental design, LLMs and LLM-powered systems were used to generate and execute the agentic workflows evaluated in this paper. We also used LLMs to enhance our text through light editing tasks such as grammar correction, and sentence restructuring.

\bibliography{architect_agent}
\bibliographystyle{icml2026}

\newpage
\appendix
\onecolumn
\section{Appendix}
\label{sec:appendix}

\subsection{Workflow Agent Runs}
\label{sec:workflowagentruns}
\FloatBarrier
\begin{figure*}[h!]
  \centering
  \includegraphics[width=\textwidth]{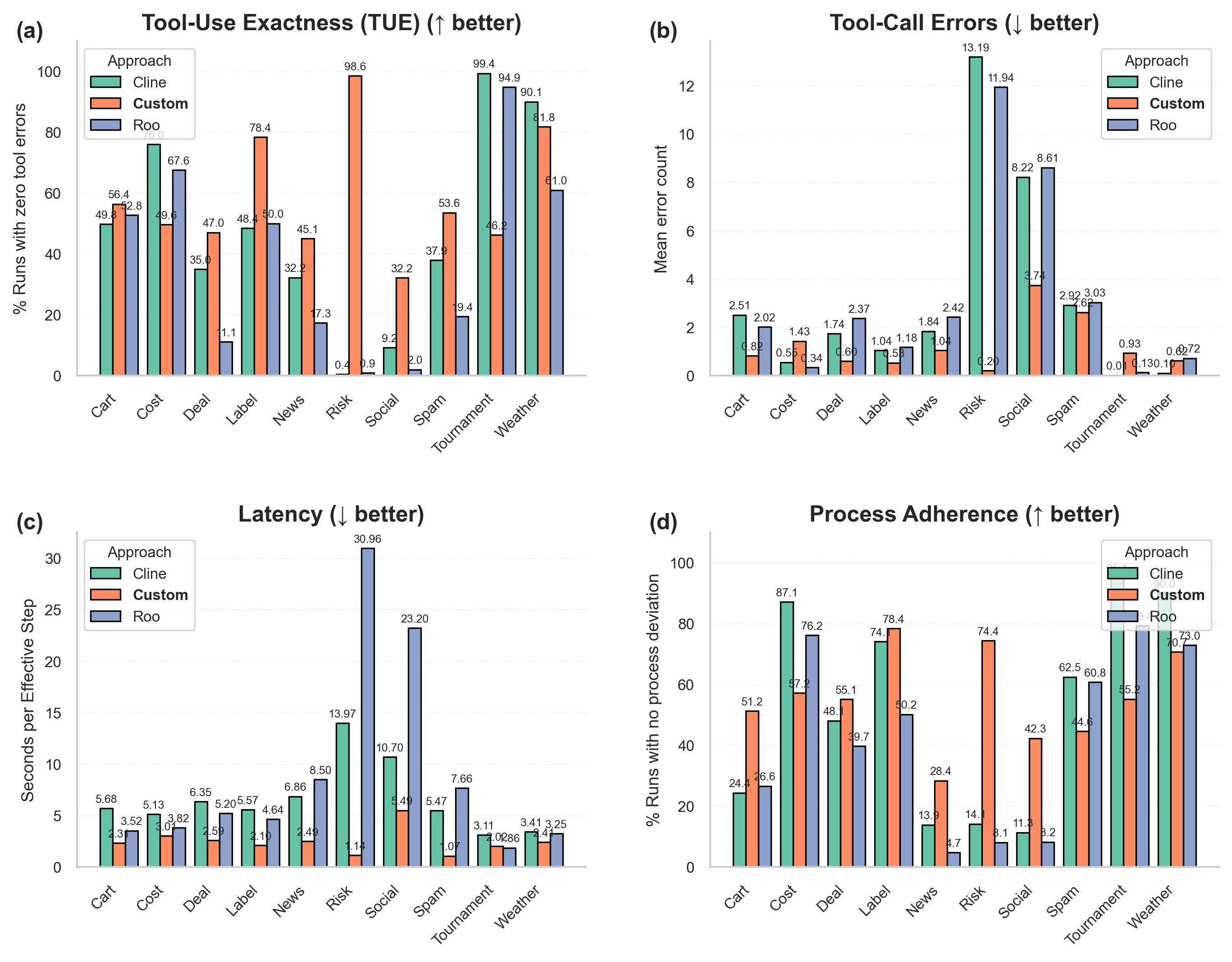}
  \caption{Per Workflow Agent Run Evaluation Metrics: Tool-Use Exactness, Tool-Call Errors, Latency and Process Adherence for the Specialist System}
  \label{fig:agent_run_evaluation_metrics}
\end{figure*}
\FloatBarrier

\newpage
\subsection{Workflow Summary Table}
\label{sec:worklowsummarytable}
\FloatBarrier
\begin{table*}[h!]
\centering
\begin{tabular}{|l|p{10.5cm}|c|c|}
\hline
\textbf{Workflow} & \textbf{Brief Overview} & \textbf{Nodes} & \textbf{Edges} \\ \hline 
Cart & Online retail order fulfillment process covering cart validation, payment authorisation, inventory reservation, shipping, delivery monitoring, and refund/reorder handling. & 52 & 60 \\ \hline
Cost & Fraud cost optimisation workflow that evaluates rules, checks fraud flags, payment status, and recoverability to compute operational costs and penalties. & 12 & 15 \\ \hline
Deal & Black Friday deal locator that routes based on user preferences (online/in-store), fetches deals, filters by relevance, and presents ranked results. & 11 & 13 \\ \hline
Labelling & User disengagement classification workflow that checks eligibility, funding status, and account closure to assign ``Disengaged'' or ``Exclude'' labels. & 13 & 16 \\ \hline
News & News article collection pipeline that queries multiple providers, validates responses, applies filters, and aggregates results with run metrics. & 34 & 38 \\ \hline
Risk & Risk and opportunity classifier that computes risk/opportunity scores, applies threshold checks, assigns priority levels, and persists classification records. & 23 & 25 \\ \hline
Social Media & Social media signal collector that fetches posts across platforms, normalises fields, filters by language/region/content type, and builds aggregated datasets. & 25 & 29 \\ \hline
Promotion & Customer promotion workflow that fetches profiles, validates eligibility, retrieves order history and loyalty points, segments customers by value tier, calculates discounts, generates recommendations, and sends notifications based on urgency. & 18 & 19 \\ \hline
Tournament & Tournament registration process that validates player eligibility, checks slot availability, submits registration, and confirms or rejects enrollment. & 14 & 17 \\ \hline
Weather & Weather forecast workflow that fetches weather data, validates retrieval, parses conditions, and displays either severe alerts or regular forecasts. & 9 & 10 \\ \hline
\end{tabular}
\caption{Summary of the ten deterministic BPMN workflows used for evaluation. Nodes include tasks, gateways, and start/end events; edges represent sequence flows between nodes.}
\label{tab:workflow_summary}
\end{table*}

\newpage
\subsection{Key Metrics by Approach}
\label{sec:keymetricsbyapproach}
\begin{figure*}[h!]
  \centering
  \includegraphics[width=\textwidth]{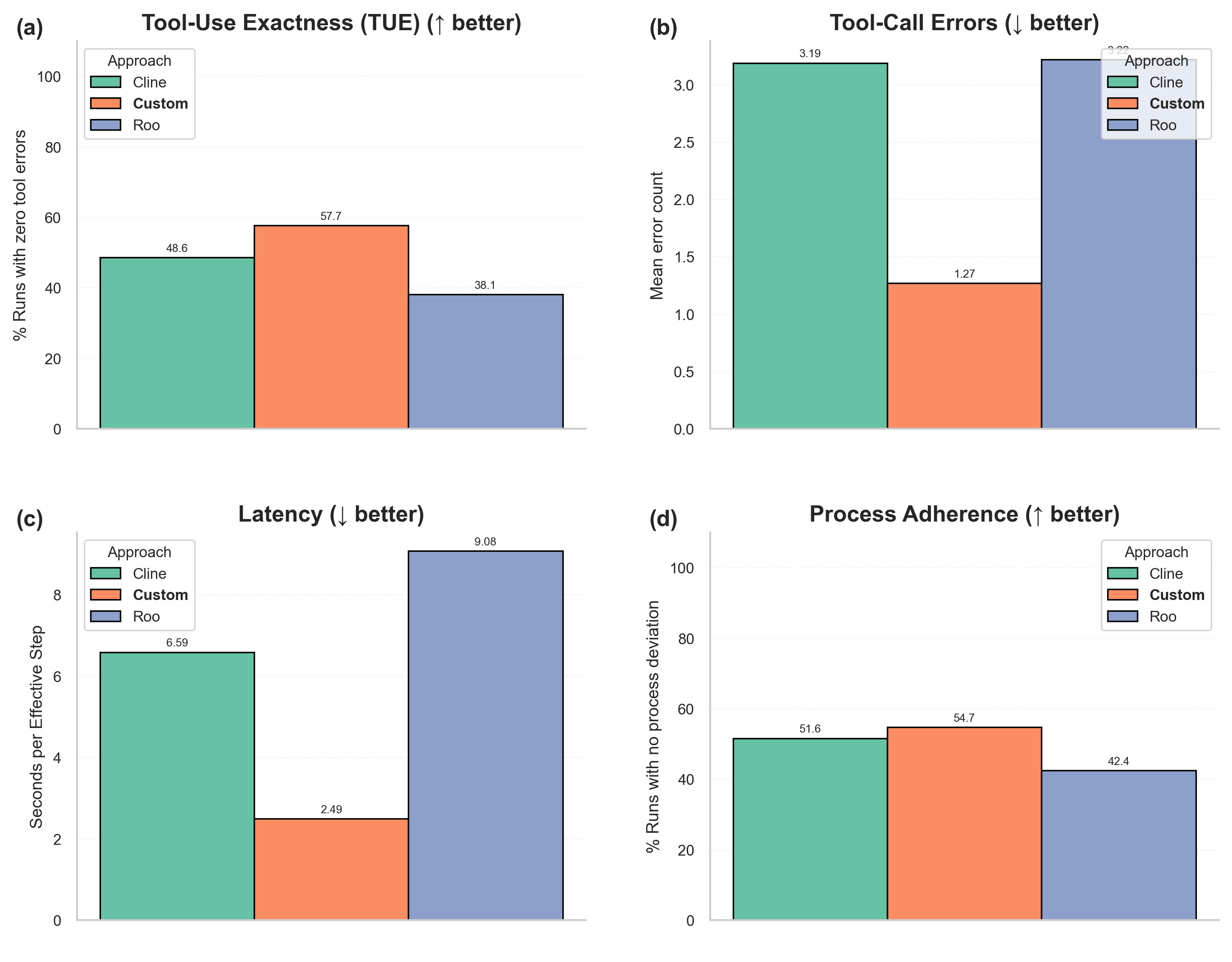}
  \caption{\textbf{Key metrics by approach.}
  Comparison of key metrics across different approaches in agent run evaluation.}
  \label{fig:appendix:key_metrics}
\end{figure*}
\FloatBarrier

\newpage
\subsection{Base Prompt for Agent Generation}
\label{sec:basepromptforagentgeneration}
\FloatBarrier
\begin{figure*}[h!]
\centering
\fbox{%
\begin{minipage}{0.95\textwidth}
\small\ttfamily\raggedright
Create a GenAI powered ReAct agent in a new folder called \textbf{\{framework\}\_agent\_experiment\_\{num\}}. Do not create a state machine, hardcoded implementation or any other implementation, your agent should be a GenAI powered ReAct agent that uses tool calls to complete the task. Do not ask any questions about what implementation to use -- you should make all of these decisions yourself.

\medskip
Your agent folder needs at least 3 files:
\begin{enumerate}
\setlength\itemsep{2pt}
    \item \textbf{agent.py} -- a FastAPI service on \texttt{localhost:7860} with two endpoints: \texttt{/chat} (accepts \texttt{conversation\_id}, \texttt{cif}, and \texttt{message}; if no message is required it can accept an empty string; returns the end result of the GenAI powered ReAct agent designed to execute the workflow) and \texttt{/get\_history} (accepts \texttt{conversation\_id}; returns the complete conversation history and all tools called by the agent including \texttt{tool\_name} \& \texttt{tool\_output}, i.e.\ \texttt{\{"conversation": [], "tool\_calls": []\}}). It should adhere to the provided API spec. Ensure this file has a \texttt{main} function that uses \texttt{uvicorn} to run the app.
    \item \textbf{tools.py} -- contains all tools for the agent.
    \item \textbf{prompts/agent\_prompt.md} -- contains the prompt for the agent.
\end{enumerate}

\medskip
Use the following to get the necessary API details for the LLM calls for your ReAct agent:

\smallskip
\texttt{openai\_base\_url = os.environ.get("OPENAI\_BASE\_URL")}\\
\texttt{openai\_api\_key\ \ = os.environ.get("OPENAI\_API\_KEY")}\\
\texttt{model\_name\ \ \ \ \ \ = os.environ.get("AGENT\_LLM")}

\medskip
Your agent should be based on the provided BPMN specification (passed in the \texttt{<BPMN>} block). You are not permitted to read any other files to complete this task. Use the venv stored at \texttt{YOUR\_VENV\_HERE}. Do not examine this venv or try to install new libraries -- if needed they will be automatically installed for you. Do not attempt to execute your code after completing it; manual testing will be performed and any errors will be passed directly to you.
\end{minipage}%
}
\caption{Base prompt provided to Roo and Cline for each agent generation attempt. The \texttt{\{num\}} placeholder was incremented per attempt, and the \texttt{<API\_SPEC>} and \texttt{<BPMN>} blocks were populated with the actual API specification and BPMN workflow for each experiment.}
\label{fig:base_prompt}
\end{figure*}

\newpage
\subsection{BPMN Specifications for Each Workflow}
\label{sec:bpmn_specifications}
\FloatBarrier
\begin{figure*}[h!]
  \centering
  \includegraphics[height=0.9\textheight,keepaspectratio]{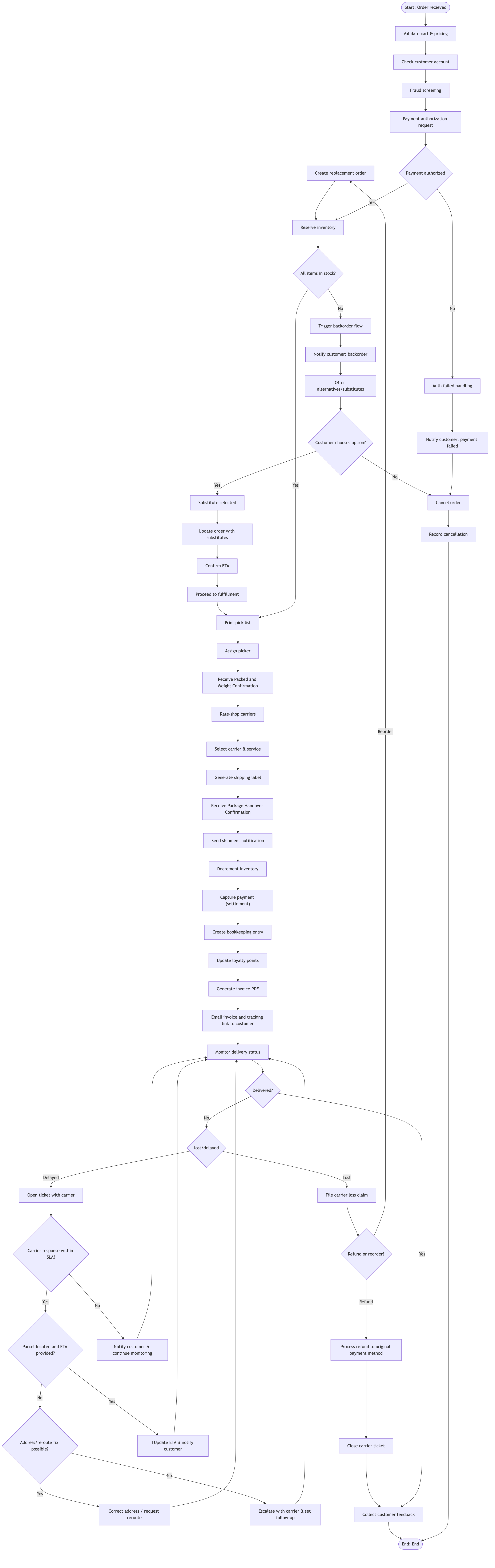}
  \caption{\textbf{Online retail cart/order fulfillment workflow.}
  The process validates the cart, authorises payment, reserves inventory, and routes to either fulfillment/shipping or cancellation/backorder handling depending on payment and stock decisions. After shipping, it monitors delivery and branches into delayed/lost handling, ending in either delivery confirmation, refund, or reorder.}
  \label{fig:appendix:cart}
\end{figure*}
\FloatBarrier

\begin{figure*}[t]
  \centering
  \includegraphics[height=0.9\textheight,keepaspectratio]{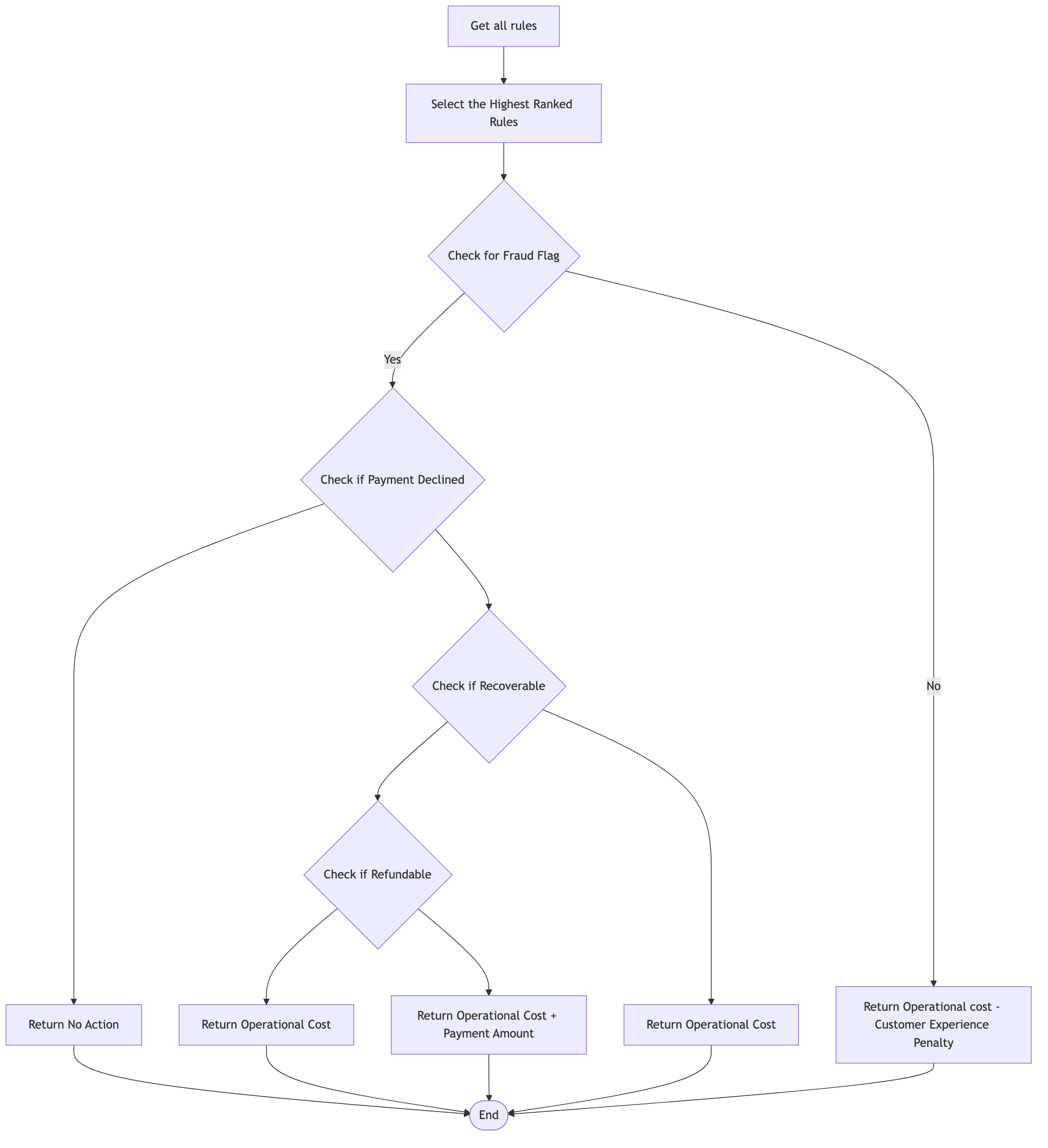}
  \caption{\textbf{Costing/estimation workflow.}
  A deterministic sequence of data-gathering and calculation tasks with decision gateways that select which cost components to apply, followed by aggregation into a final cost/price output.}
  \label{fig:appendix:cost}
\end{figure*}
\FloatBarrier

\begin{figure*}[t]
  \centering
  \includegraphics[height=0.95\textheight,keepaspectratio]{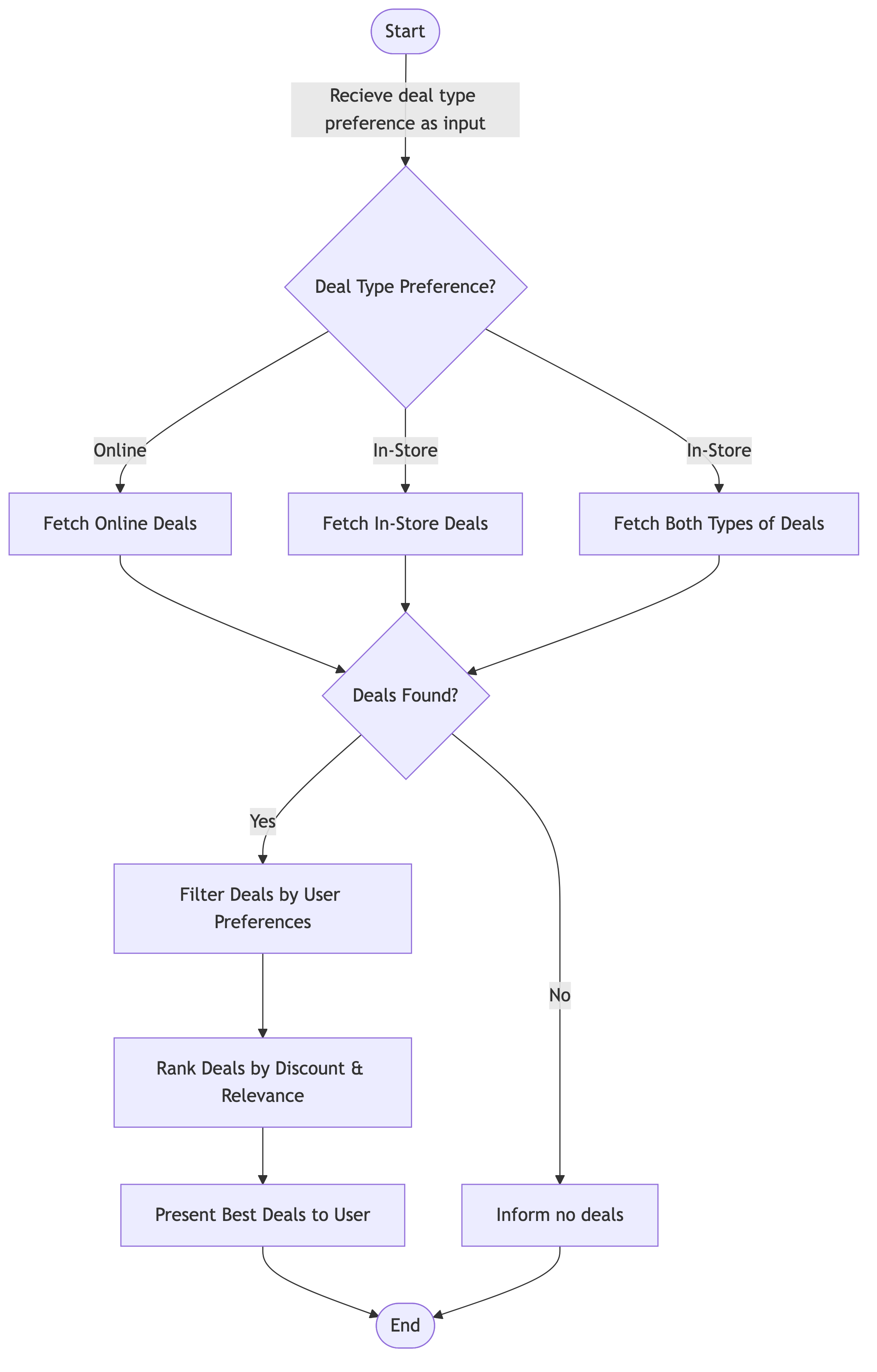}
  \caption{\textbf{Deal assessment workflow.}
  The workflow evaluates a candidate deal through staged checks and scoring steps, using decision gateways to route to accept/reject/escalate outcomes based on thresholds and validation checks.}
  \label{fig:appendix:deal}
\end{figure*}
\FloatBarrier

\begin{figure*}[t]
  \centering
  \includegraphics[height=0.95\textheight,keepaspectratio]{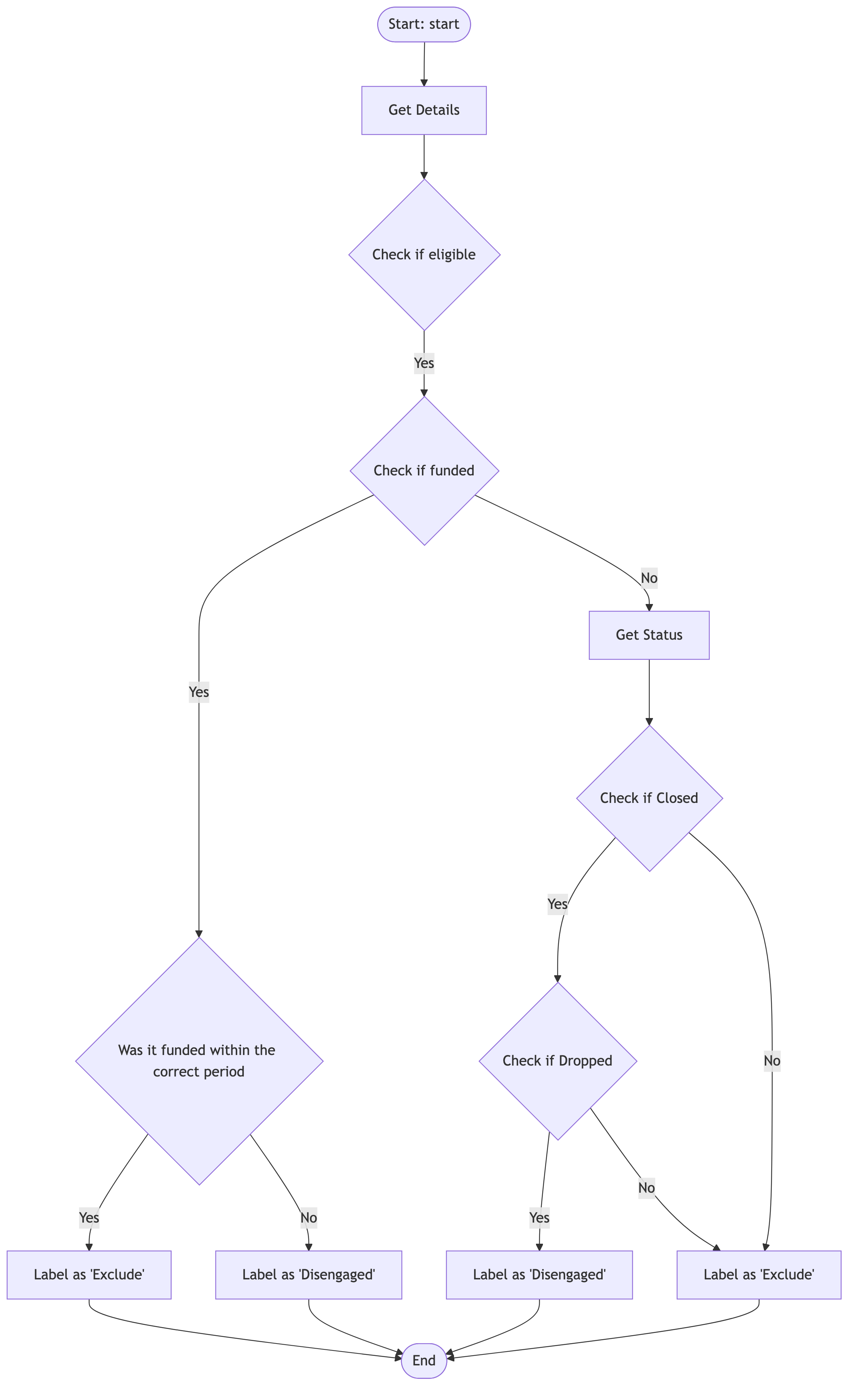}
  \caption{\textbf{Labelling/classification workflow.}
  A rules/criteria-driven labelling pipeline that applies sequential checks and uses gateways to assign labels and/or exclude items, terminating once a label decision is reached.}
  \label{fig:appendix:labelling}
\end{figure*}
\FloatBarrier

\begin{figure*}[t]
  \centering
  \includegraphics[height=0.95\textheight,keepaspectratio]{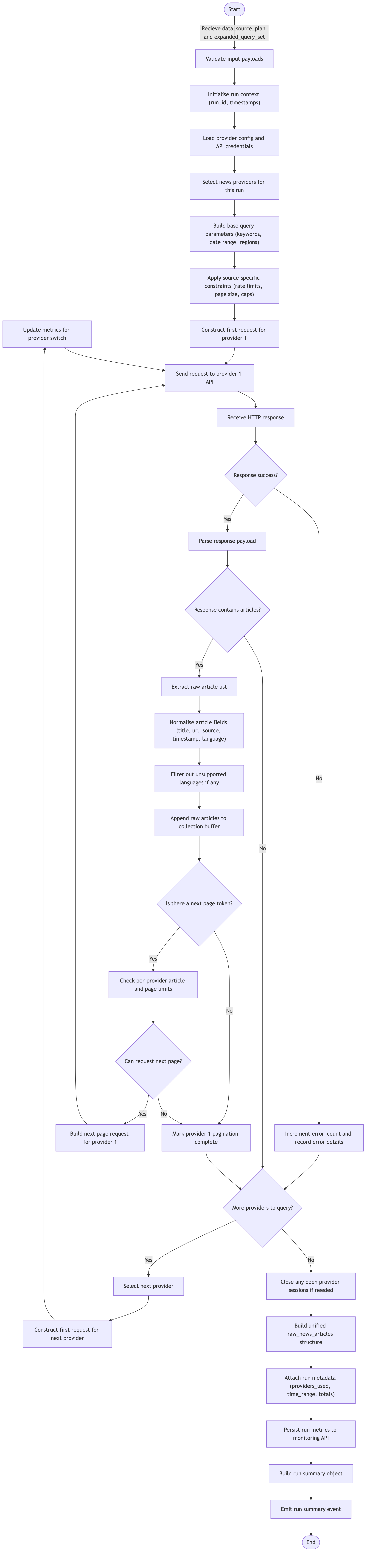}
  \caption{\textbf{News processing workflow.}
  The process ingests an item, performs validation and enrichment steps, then routes through conditional branches (e.g., eligibility/quality thresholds) to produce a final categorisation/output or a discard/escalation outcome.}
  \label{fig:appendix:news}
\end{figure*}
\FloatBarrier

\begin{figure*}[t]
  \centering
  \includegraphics[width=0.24\textwidth,keepaspectratio]{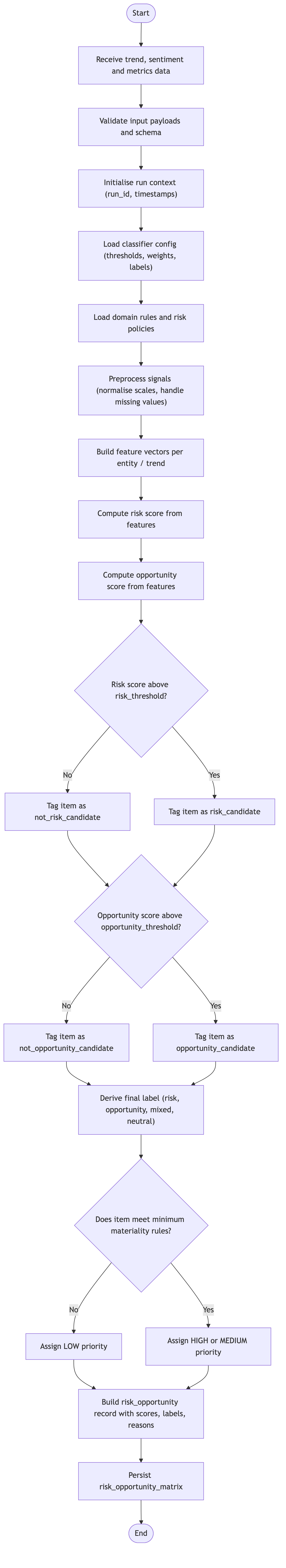}
  \caption{\textbf{Risk \& opportunity classifier workflow.}
  The workflow computes risk/opportunity signals, derives a final label (risk/opportunity/mixed/neutral), then routes by a decision gateway into different priority/handling paths (e.g., high/medium vs low) before completing.}
  \label{fig:appendix:risk}
\end{figure*}
\FloatBarrier

\begin{figure*}[t]
  \centering
  \vspace{1.5cm}
  \includegraphics[width=0.233\textwidth,keepaspectratio]{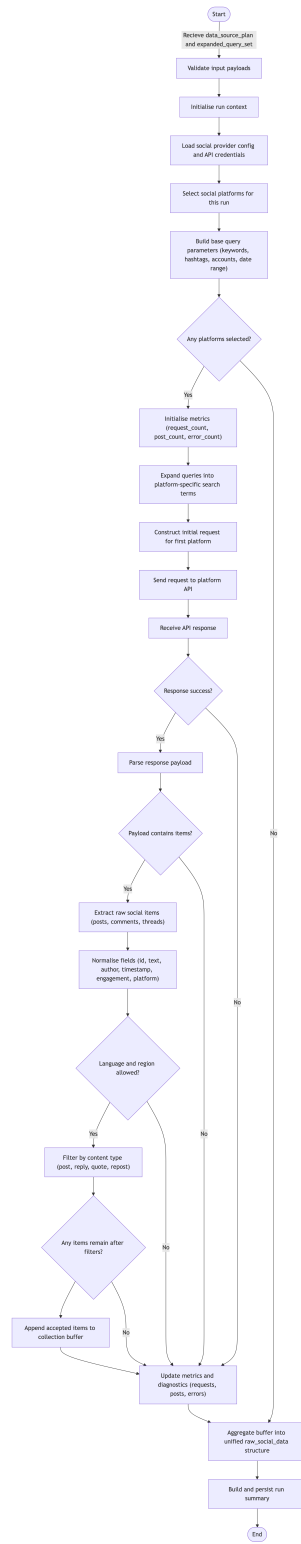}
  \caption{\textbf{Social media workflow.}
  A multi-step pipeline that performs content/user/context checks and then uses decision gateways to choose an action path (e.g., allow, flag, or escalate), ending once an action outcome is produced.}
  \label{fig:appendix:socialmedia}
\end{figure*}
\FloatBarrier

\begin{figure*}[t]
  \centering
  \includegraphics[height=0.95\textheight,keepaspectratio]{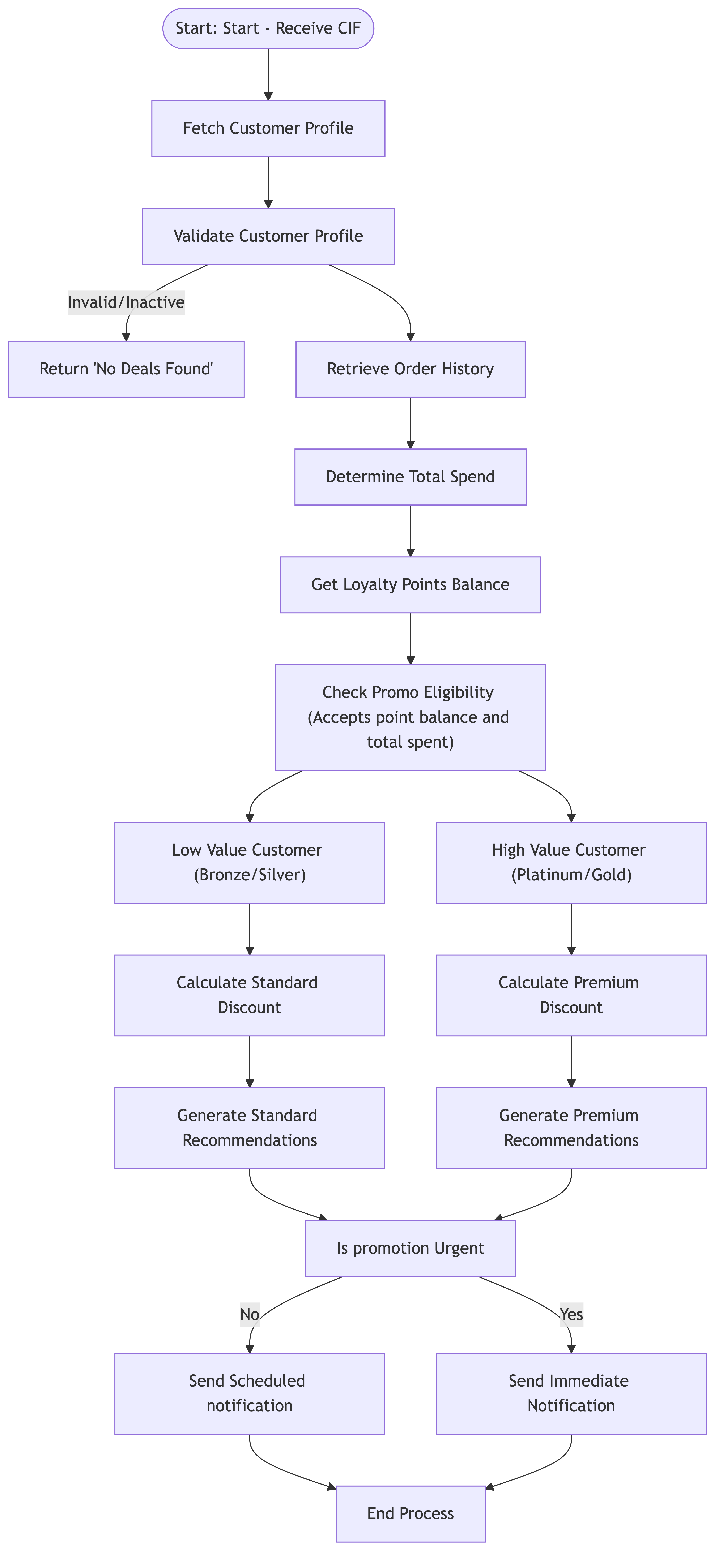}
  \caption{\textbf{Customer promotion workflow.}
  The process fetches and validates customer profiles, retrieves order history and loyalty points, checks promotion eligibility, segments customers into value tiers (high/low), calculates tier-appropriate discounts, generates recommendations, and routes notifications based on promotion urgency.}
  \label{fig:appendix:promotion}
\end{figure*}
\FloatBarrier

\begin{figure*}[t]
  \centering
  \includegraphics[height=0.95\textheight,keepaspectratio]{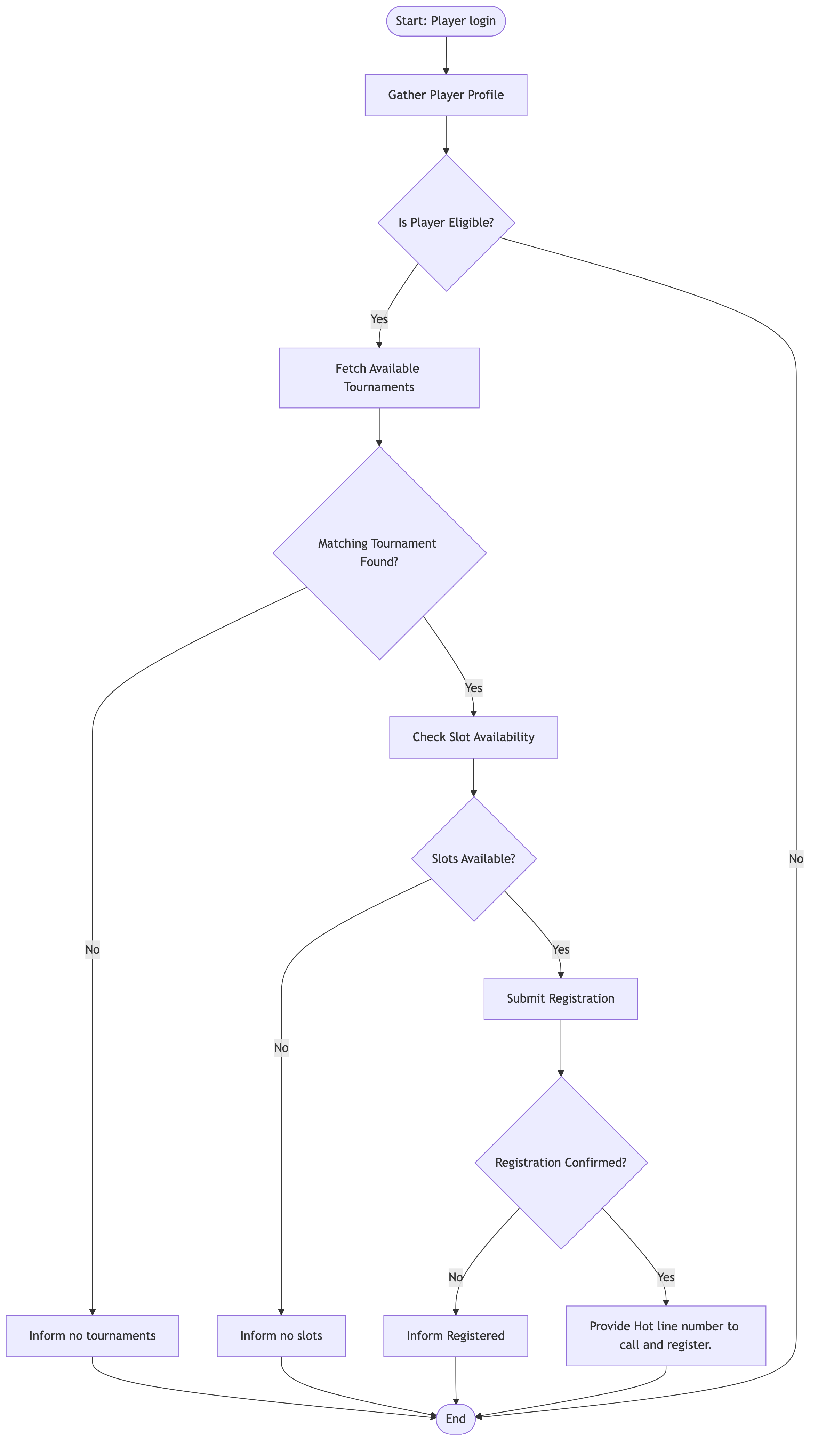}
  \caption{\textbf{Tournament/bracket workflow.}
  The workflow advances items/participants through staged rounds, using decision gateways to determine advancement and termination once a final winner/outcome is determined.}
  \label{fig:appendix:tournament}
\end{figure*}
\FloatBarrier

\begin{figure*}[t]
  \centering
  \includegraphics[height=0.95\textheight,keepaspectratio]{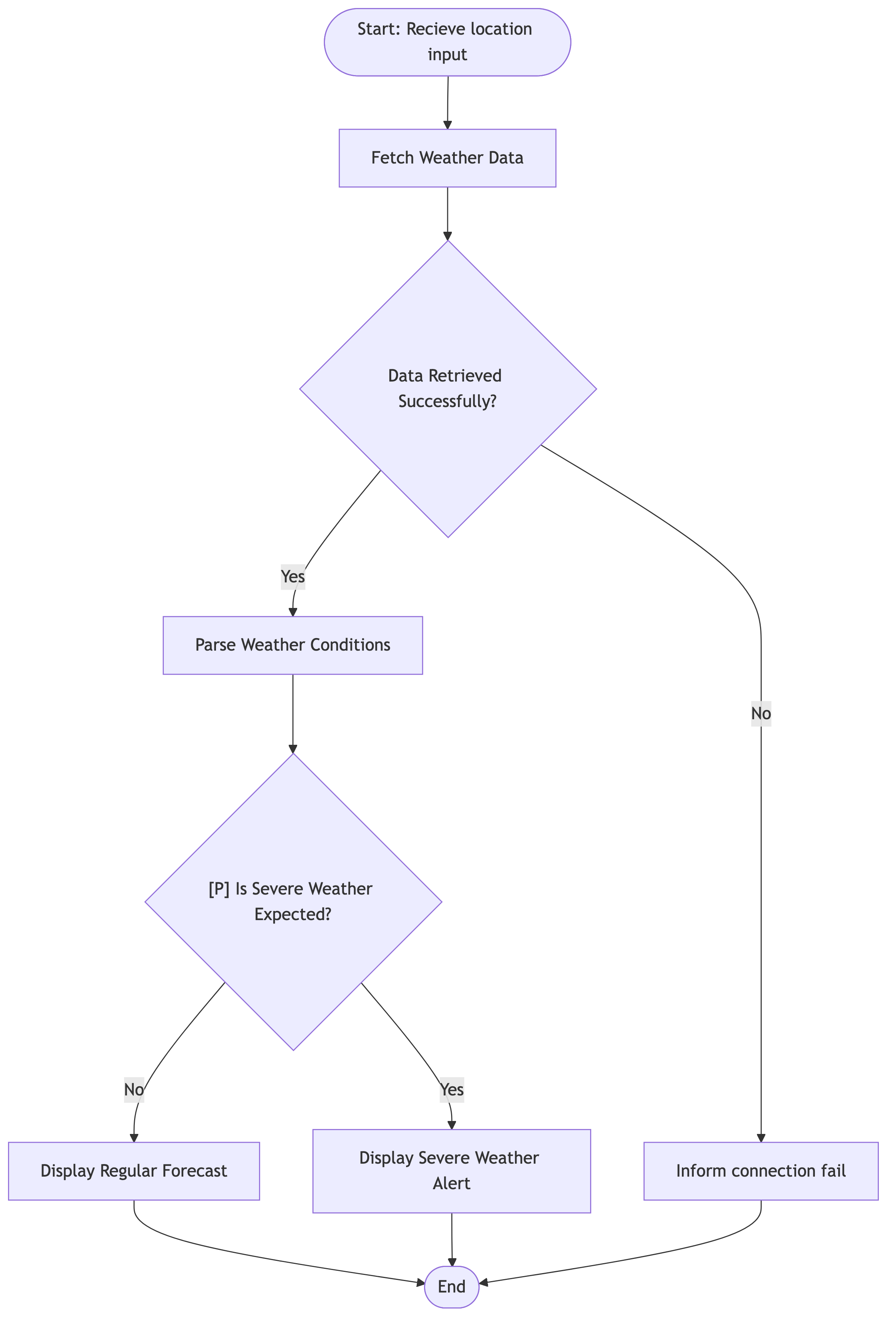}
  \caption{\textbf{Weather workflow.}
  A deterministic sequence of data retrieval and transformation steps with conditional branches (e.g., data availability/threshold checks), culminating in a final forecast/decision output.}
  \label{fig:appendix:weather}
\end{figure*}
\FloatBarrier

\end{document}